\documentclass[iop,apj,tighten]{emulateapj}
\usepackage{apjfonts} %If missing fonts, comment out this or google apjfonts to download
\usepackage{amsmath,amstext}
\usepackage[breaklinks,colorlinks,citecolor=blue,linkcolor=blue]{hyperref}

\providecommand{\araa}[0]{Ann.\ Rev. Astron. Astroph. }
\providecommand{\aap}[0]{Astron. Astrophys. }

\providecommand{\aj}[0]{Astron. J. }

\providecommand{\apj}[0]{Astrophys. J.~}
\providecommand{\apjl}[0]{Astrophys. J. Lett. }

\providecommand{\mnras}[0]{Mon. Not. Roy. Astron. Soc. }

\providecommand{\prd}{Phys. Rev. D. }

\newcommand{\SMBH}{\bullet}
\newcommand{\Msun}{M_{\odot}}

\newcommand\lsim{\mathrel{\rlap{\lower4pt\hbox{\hskip1pt$\sim$}}
        \raise1pt\hbox{$<$}}}
\newcommand\gsim{\mathrel{\rlap{\lower4pt\hbox{\hskip1pt$\sim$}}
        \raise1pt\hbox{$>$}}}

\shorttitle{Stellar-mass Binary Black Hole Mergers in AGN}
\shortauthors{Bartos, Kocsis, Haiman, \& M\'arka}

\begin{document}

\title{Rapid and Bright Stellar-mass Binary Black Hole Mergers in Active Galactic Nuclei}

\author{Imre Bartos\altaffilmark{1}, Bence Kocsis\altaffilmark{2}, Zolt\'an Haiman\altaffilmark{3}, and Szabolcs M\'arka\altaffilmark{1}}
\affil{$^1$ Department of Physics, Columbia University, 550 W120th Str., New York, NY 10027, USA}
\affil{$^2$Institute of Physics, E\"otv\"os University, P\'azm\'any P. s. 1/A, Budapest, 1117, Hungary}
\affil{$^3$Department of Astronomy, Columbia University, 550 W120th Str., New York, NY 10027, USA}

%%%%%%%%%%%%%%%%%%%%%%%%%%%%%%%%%%%%%%%%%%%%%%%%%%%%%%%%%%%%%%%%%%%%%%

\begin{abstract}
The Laser Interferometer Gravitational-Wave Observatory, LIGO, found direct evidence for double black hole binaries emitting gravitational waves. Galactic nuclei are expected to harbor the densest population of stellar-mass black holes. A significant fraction ($\sim30\%$) of these black holes can reside in binaries. We examine the fate of the black hole binaries in active galactic nuclei, which get trapped in the inner region of the accretion disk around the central supermassive black hole. We show that binary black holes can migrate into and then rapidly merge within the disk well within a Salpeter time. The binaries may also accrete a significant amount of gas from the disk, well above the Eddington rate. This could lead to detectable X-ray or gamma-ray emission, but would require hyper-Eddington accretion with a few percent radiative efficiency, comparable to thin disks. We discuss implications for gravitational wave observations and black hole population studies. We estimate that Advanced LIGO may detect $\sim20$ such, gas-induced binary mergers per year.
\end{abstract}


\maketitle

%%%%%%%%%%%%%%%%%%%%%%%%%%%%%%%%%%%%%%%%%%%%%%%%%%%%%%%%%%%%%%%%%%%%%%

\section{Introduction}

The study of stellar-mass binary black holes (BBHs) is greatly limited by their expected lack of electromagnetic radiation. However, the merger of such systems produces luminous radiation in gravitational waves, making them prime targets for Earth based detectors, including Advanced LIGO \citep{2015CQGra..32g4001L}, Advanced Virgo \citep{2015CQGra..32b4001A} and KAGRA \citep{2013PhRvD..88d3007A}. The detection of BBH mergers present a unique opportunity to probe general relativity in the strong field regime \citep{2006LRR.....9....3W}, to study the formation, evolution and environment of black holes \citep{2015arXiv151204897A}, and may give us observational probes of the black hole mass function and dynamics in galactic nuclei and globular clusters \citep{2016arXiv160202444R,2016arXiv160202809O}. The recent discovery of a BBH merger by Advanced LIGO \citep{PhysRevLett.116.061102} has made the prospects of studying these systems particularly interesting.

A BBH system can form in isolated stellar binaries \citep{2002ApJ...572..407B,2012ApJ...759...52D,Marchant16,Mandel16} and in dynamical interactions in dense stellar systems including globular clusters and galactic nuclei \citep{2002ApJ...576..899P,millerhamilton02,2004Natur.428..724P,2009MNRAS.395.2127O,2012PhRvD..85l3005K,2013ApJ...773..187N,Antonini2014ApJ...781...45A,Antognini2014MNRAS.439.1079A,2015ApJ...800....9M}. The binary subsequently loses energy and angular momentum and eventually merges via gravitational wave radiation reaction. In the presence of stars or gas near the binary, its dynamical evolution may also be affected by energy and angular momentum exchange during stellar encounters \citep{2014ApJ...782..101P} or interactions with the gaseous medium. BBHs near supermassive black holes (SMBHs) are also subject to secular Kozai-Lidov processes, which can alter the inclination and eccentricity of their orbit, potentially decreasing the merger time \citep{2012ApJ...757...27A,2016MNRAS.460.3494S}.

\emph{Active galactic nuclei} (AGNs) represent a conceivable location where BBHs may be commonly embedded in dense gaseous environments. In AGNs, infalling gas cools radiatively and forms a cold accretion disk around the central SMBH, which powers the observed highly luminous emission. Galactic nuclei are also expected to harbor large populations of stellar-mass black holes, which sink towards the central SMBH due to dynamical friction on stars \citep{1993ApJ...408..496M,2004cbhg.symp..138R,2009MNRAS.395.2127O,2013PhRvL.110v1102B}. Heavier objects sink faster. Nearby globular clusters, essentially behaving as even heavier objects, can also migrate into the galactic nucleus  \citep{TremaineOstrikerSpitzer1975}e, carrying black holes with them. Further, massive stars can form near galactic centers, e.g., in AGN disks \citep{GoodmanTan2004,Levin2007,McKernan+2012,2016arXiv160204226S}, providing an additional source of black holes to the nucleus.

A significant fraction of high-stellar-mass objects reside in binaries. As much as 70\% of massive stars are observed to have a companion \citep{2007ApJ...670..747K,2012Sci...337..444S}. In the Milky Way nucleus, the binary fraction of massive stars is observationally estimated to be $\sim 30\%$ \citep{2014ApJ...782..101P}. BBHs can form via \emph{isolated stellar binary systems} \citep{2016ApJ...818L..22A,2016MNRAS.458.2634M} if the two massive stars undergo core collapse, creating black holes without disrupting the binary or overly widening the orbit. Alternatively, BBHs can be formed \emph{dynamically} through the chance encounters of black holes in dense stellar environments, such as galactic nuclei \citep{2009MNRAS.395.2127O} or globular clusters \citep{2006ApJ...648..411K,2015PhRvL.115e1101R}.

In this paper, we study the evolution of a BBH population in a newly activated AGN, and its observational consequences. Galactic centers can experience a large influx of gas, e.g., in galactic mergers, or as a result of secular instabilities.
The newly formed accretion disk will interact with nearby BBHs, accelerating their merger. Mergers within the disk can also be accompanied by amplified accretion by the BBH, producing an electromagnetic counterpart to the gravitational wave signal.

In the following, we first show that BBHs within the disk can rapidly ($\lesssim1$\,Myr) merge by interacting with the gas in a typical AGN disk (Section \ref{section:dynamicalfriction}). We then show that a significant fraction of the BBHs in the nucleus will rapidly ($\lesssim10$\,Myr) align their orbital axes with the disk due to loss of momentum upon disk crossings, increasing the number of BBHs that reside within the disk (Section \ref{section:orbitalalignment}). In Section \ref{section:RATE}, we discuss the expected rate of BBH mergers within AGN disks within the detectable range of Earth based gravitational wave detectors. We examine the production and detectability of X-ray and gamma-ray emission due to amplified accretion onto the binary from the disk (Section \ref{section:EM}). We present our conclusions in Section \ref{section:conclusion}.

\section{Rapid merger via Gaseous Torques}
\label{section:dynamicalfriction}

In the absence of a circumbinary medium and encounters with other stars, the orbital evolution of BBHs is dictated by gravitational wave emission.
Gravitational wave emission  diminishes rapidly with separation, and therefore a sufficiently wide binary spends most of its lifetime close to its initial separation following its formation. For instance, a circular 10+10\,M$_{\odot}$ BBH with a 13\,hour period merges in a Hubble time, but one with a $1$\,hour period merges in 10\,Myr.

The pace of the orbital decay can greatly increase within a gaseous environment, due to tidal and viscous angular momentum exchange between the binary and the surrounding gas \citep{2008ApJ...679L..33K,2010MNRAS.402.1758S,2011ApJ...726...28B}. This interaction (either ``dynamical friction'' or resonant angular momentum transport) can harden the binary from wider separations to the point where gravitational-wave emission can take over, resulting in a rapid merger.

When gas is delivered to the nucleus and activates the central AGN, angular momentum conservation and radiative cooling leads to the formation of a thin accretion disk around the central SMBH.  Stellar-mass BBHs may then be located in, or migrate into, a gaseous environment in the newly formed AGN disk. Since black holes are more massive than typical stars, black holes which formed in the local neighborhood of the galactic center sink to the center due to two-body encounters in the nuclear stellar cluster \citep{2000ApJ...545..847M,2009MNRAS.395.2127O}. Black holes may also be delivered to this region by star clusters which sink to the center due to dynamical friction against dark matter, gas, and stars in the galactic disk and halo. Massive stars are also believed to form in the outer, unstable regions of
an AGN disk  \citep{GoodmanTan2004,Levin2007,2011PhRvD..84b4032K,McKernan+2012,TQM05}.  As a result, black hole binaries may also form in these regions, from massive stellar binaries which were created in the disk and/or grew massive through accretion from the disk (see also Ref.~\citep{2016arXiv160204226S}).

Alternatively, black hole binaries may form in the vicinity of the SMBH from binary stars that were scattered into the galactic center and got captured through the Hills mechanism \citep{1990AJ.....99..979H}. In this process, the binary gets scattered on a nearly radial orbit toward the SMBH, interacts with the SMBH, and the less massive component is ejected as a hypervelocity star, while the massive component gets captured on a close orbit representing the S-stars in the Galactic center. Such stars may be dragged into the accretion disk through hydrodynamic drag and vector resonant relaxation and eventually form black holes there \citep{1996NewA....1..149R,2005A&A...433..405S}. Single objects may form close binaries after dynamical multibody interactions or gravitational wave capture. Binaries initially on wider orbits in the nuclear stellar cluster may also be dragged into the accretion disk if they get scattered onto orbits that cross the accretion disk \citep{1991MNRAS.250..505S,2001A&A...376..686K}. Triple disruption may also efficiently bring binaries into the nucleus \citep{2009ApJ...690..795P,2011arXiv1109.2284G}.

In the following we characterize the merger time scale of a stellar-mass BBHs, embedded in an accretion disk around a SMBH. We consider a geometrically thin, optically thick, radiatively efficient, steady-state accretion disk \citep{1973A&A....24..337S}, which is expected in AGNs \citep{2011PhRvD..84b4032K}. We adopt the viscosity parameter $\alpha = 0.3$ \citep{2007MNRAS.376.1740K}. We assume a central SMBH with mass $M_{\bullet} = 10^{6}$\,M$_{\odot}$, comparable to the SMBH mass within the Milky Way, accreting at $\dot{m}_{\bullet} = \dot{M}_{\bullet} / \dot{M}_{\bullet,\rm Edd} = 0.1$ relative to the Eddington rate $\dot{M}_{\bullet,\rm Edd}$ \citep{2011PhRvD..84b4032K}, with radiation efficiency $\epsilon = L_{\bullet,\rm Edd} / \dot{M}_{\bullet,\rm Edd}c^2= 0.1$. These parameters determine the disk surface density $\Sigma(R)$, scale height $H(R)$, isothermal sound speed $c_s(R)$, and midplane temperature, $T$ (see Eqs.~26, 27, 28, 29, and 37 in \citealt{2011PhRvD..84b4032K} with $b=0$; and Eq.~11 in \citealt{2009ApJ...700.1952H} where we fix a typo of order unity, we multiply the expression by $(3/4)^{1/4}$, respectively).
At the characteristic distance from the SMBH $R_{\rm disk}=10^{-2}$pc,
we find\footnote{These asymptotic radial scaling relations are only valid in the gas pressure dominated regime.}
\begin{eqnarray}
\Sigma &\approx& 2000\,\mbox{g}\,\mbox{cm}^{-2}\,(R/10^{-2}\mbox{pc})^{-3/5}(M_{\bullet}/10^6 M_\odot)^{4/5}\\
H &\approx& 10^{14}\,\mbox{cm}\,(R/10^{-2}\mbox{pc})^{21/20}(M_{\bullet}/10^6 M_\odot)^{-3/20}\\
c_s &\approx& 2\,\mbox{km}\,\mbox{s}^{-1}(R/10^{-2}\mbox{pc})^{-9/20}(M_{\bullet}/10^6 M_\odot)^{3/2}\\
T &\approx& 800\,\mbox{K}\,(R/10^{-2}\mbox{pc})^{-9/10}(M_{\bullet}/10^6 M_\odot)^{7/10}
\end{eqnarray}

\subsection{Drag force}

%If the orbital separation of the stellar-mass BBH is smaller than its Hill radius, then it can be surrounded by gas, bound to the BBH.  This
Gas surrounding the BBH can provide a drag on the binary and facilitate its merger.  The geometry of this gas is poorly understood, and we here employ two different simplified models to estimate the drag experienced by the binary.

First, we follow the semianalytic results of \citet{2007ApJ...665..432K,2008ApJ...679L..33K}. Kim et al. compute the drag force of an equal-mass binary perturber, with a circular orbit, embedded in a uniform and isothermal gaseous background medium, as a function of the perturbers' Mach number $\mathcal{M}\equiv v_{p}/c_s$. Here $v_{p}$ is the orbital velocity of the perturber, measured from the binary's center of mass.  The perturbers create density wakes within the gaseous medium, which act as a drag force, analogous to dynamical friction \citep{1999ApJ...513..252O}. Because of the circular motion, the wakes from each perturber interacts with the other perturber, reducing the total drag force, making the interaction important in estimating the binary evolution.  We assume that the center-of-mass (CoM) of the binary is orbiting the central SMBH at approximately the same speed as the gas bound to the binary. We also assume that the gas cloud around the BBH has no net rotation around the binary's COM.  The effect of the finite thickness of the accretion disk is expected to be small for our typical model parameters. At a distance $R = 0.01$\,pc from the central SMBH, we find a disk scale height $H\sim 10^{14}$\,cm, which is much greater than the characteristic distance $d_s = c_s t_{\rm orb}$ sound travels within the gas during the orbital period $t_{\rm orb}$. \citet{2008ApJ...679L..33K} find that the gravitational drag force converges to its steady-state value within $\sim 1$ orbital period, indicating that dynamical friction is mostly the effect of gas perturbations in the vicinity of the binary.

\begin{figure}
\begin{center}
\resizebox{0.49\textwidth}{!}{\includegraphics{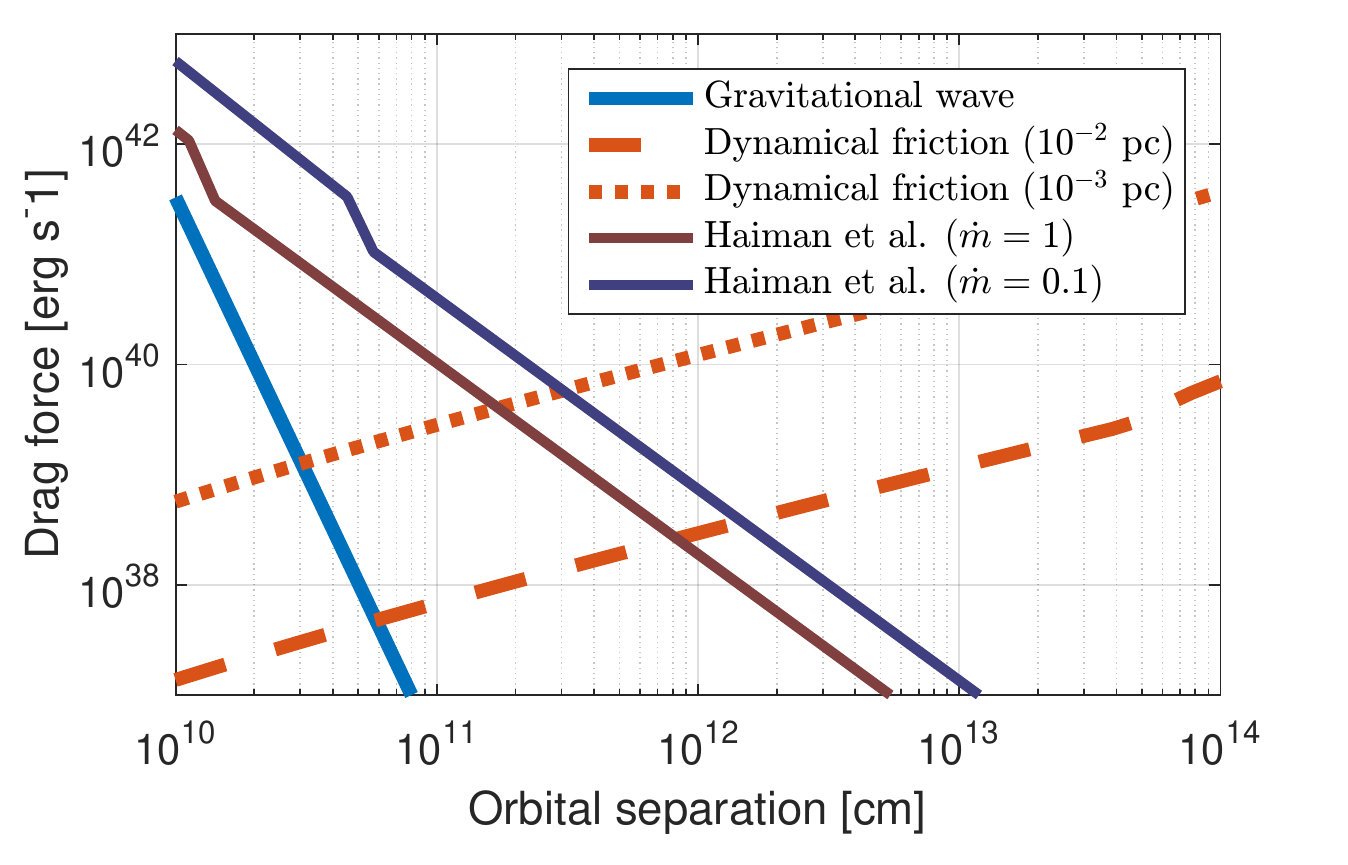}}
\end{center}
\caption{Drag force (energy loss rate) for binary black holes within a
  gas disk as a function of orbital separation due to the emission of
  gravitational waves (solid) and dynamical friction at different
  orbital radii (see legend). The parameters adopted are $M_1=M_2=10\,\Msun$, $M_{\bullet}=10^6\,\Msun$.}
\label{fig:dragforce}
\end{figure}

To calculate the drag force, we take an equal-mass binary with total mass $M_{\rm tot}=M_1+M_2=20\,$M$_{\odot}$ in a gaseous environment with density $\rho_{0} = \Sigma / 2 H$, i.e. the background density of the AGN disk around the binary.  Note that we obtain similar results for other binary masses. We calculate the total drag force on the binary components using the approximations for the drag force from their own wakes, and from the wake of the companion,  given in Eq. (14) of
\citet{2007ApJ...665..432K} and in Eq. (5) of \citet{2008ApJ...679L..33K}, respectively, as a function of the orbiting black holes' Mach number. Beyond $\mathcal{M} = 8$ for which the numerical fits in \citet{2007ApJ...665..432K,2008ApJ...679L..33K} are uncertain, we assume that the drag force is $F_{\rm df}\mathcal{M}^{2} = const.$, as expected in the $\mathcal{M}\gg 1$ limit \citep{1999ApJ...513..252O}. To obtain the drag force, we define the
characteristic size $r_{\rm min}$ \citep{2007ApJ...665..432K} of the perturbers to be the innermost stable circular orbit of the black
holes, assuming no rotation. The drag force depends logarithmically on
the choice of $r_{\rm min}$. The obtained drag force is shown for
$R=10^{-2}$\,pc in Fig. \ref{fig:dragforce}. We see that the drag
force slowly but monotonically decreases with decreasing separation,
as $\mathcal{M}>1$ for all orbital separations shown here.

An alternative, and perhaps more realistic model is that the gas bound
to the BBH has net angular momentum around the CoM of the BBH, and
forms a circumbinary mini-accretion disk.  Such a minidisk can be fed
through streams of gas from the background AGN disk, and has been seen
to form in hydrodynamical simulations for a low-mass binary perturber
\citep{1999ApJ...526.1001L, Dorazio+2016}.  In this case, the background AGN disk can
again be ignored, and the orbital decay of the BBH arises from the
torques due to the density perturbations it creates in the minidisk
\citep{1980ApJ...241..425G,2009MNRAS.393.1423C,2012A&A...545A.127R,2013MNRAS.436.2997D}.  Note that unlike in the case of the gas drag in a non-rotating cloud, the orbital inspiral caused by a minidisk accelerates as the binary separation decreases.

For sufficiently small binary separations, the effective drag force in either case is dominated by gravitational-wave emission. We approximate the energy loss due to gravitational waves in the non-relativistic limit assuming circular orbits \citep{1964PhRv..136.1224P}. We find that the resulting gravitational wave drag force dominates for binary separations below $r\sim10^{10-11}$\,cm (Fig. \ref{fig:dragforce}).

\subsection{Merger time scale}

We saw above that the drag force due to dynamical friction decreases, while the drag force due to gravitational waves quickly increases with decreasing orbital separation. As a result, orbital decay
will be the slowest at the separation where the gravitational wave emission takes over. This central separation will dominate the merger time, see  Fig. \ref{fig:mergertime}.
This transition at a separation around $10^{11}$\,cm, is only weakly dependent on $R$.
Most importantly, we find that the total merger time from an initial separation $r\leq 10^{14}$\,cm is $t_{\rm merge} < 1$\,Myr for $R\lesssim 1$\,pc. This rapid merger time is much faster than expected purely from gravitational wave emission, and is much shorter than the typical lifetime ($10^{7-8}$ yr) of bright AGN accretion disks, or binaries outside of gaseous environments.

To estimate merger times in the alternative scenario, with the gas torques arising from a thin minidisk, we use the models in \citet{2009ApJ...700.1952H}. They obtained the orbital decay of a SMBH binary by calculating binary-disk interactions, using simple thin disk models \citep{1995MNRAS.277..758S}. Adopting their fiducial disk parameters, we scale down their results to stellar mass BBHs with our model parameters. To adopt their model to our case, we need to define the accretion rate $\dot{M}_1$ (and $\dot{M}_2$) onto the binary. We use the Eddington rate for the calculation; the actual rate can be super-Eddington, given the large disk density in the vicinity of the binary for the $R$ range considered here, which would likely lead to more rapid orbital decay.  For the binary separations considered here, we are in the ``outer disk" regime, as defined in Eq. (14) of \citet{2009ApJ...700.1952H}, and we thus use Eq. (26c) of \citet{2009ApJ...700.1952H}.  We find that the characteristic timescale of the merger is $t\sim3\times10^4$\,yr for $r=10^{14}$\,cm (see  Fig. \ref{fig:mergertime}). This result also indicates that the binary merger within the inner AGN disk is expected to be very rapid.

\subsection{Region of validity}
\label{sec:regionofvalidity}

The above results are subject to the validity of the assumptions we made. First, the above thin accretion disk model for the SMBH is only valid if the
self gravity of the disk is negligible. This is valid for radii at which the disk's Toomre Q-parameter is $Q=c_{s}\Omega / (\pi G \Sigma)\gsim 1$, where $\Omega$ is the angular frequency of the gas around the SMBH. We find that $Q=1$ around $R\sim10^{-2}$pc. Beyond $10^{-2}$pc, the disk can be subject to fragmentation. The structure of the accretion disk is not well understood beyond this point. One possibility is that additional source of heating from star formation stabilizes the disk with $Q=1$ in the outer region \citep{2003MNRAS.341..501S,TQM05} but we conservatively do not extend our calculations to this regime (see \citealt{2016arXiv160204226S} for a discussion of forming BBHs in this scenario).

% IONIZATION (aka EVAPORATION)

\begin{figure}
\begin{center}
\resizebox{0.49\textwidth}{!}{\includegraphics{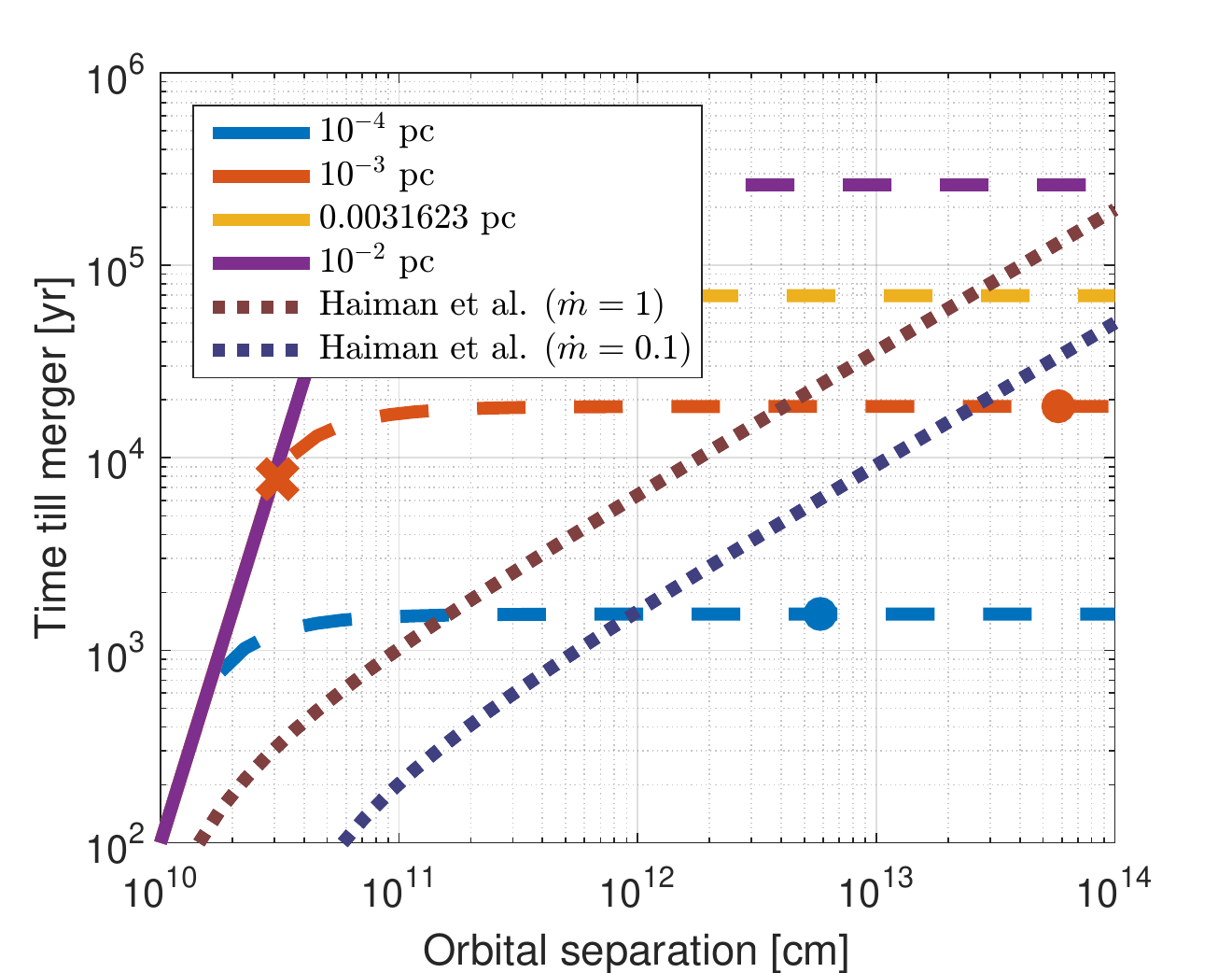}}
\end{center}
\caption{Time until merger for binary black holes as a function of their orbital separation, for different distances from the SMBH (see
  legend). The solid, dashed, and dotted curves indicate the dominant force behind orbital decay (gravitational waves, dynamical friction
  in a nonrotating cloud, or torques in a circumbinary minidisk, respectively). The circumbinary minidisks are assumed to have accretion rates $\dot{m}=0.1$ and $\dot{m}=1$ in Eddington units for the stellar-mass BHs.  Solid circles and crosses indicate the separations that are equal to the Hill radius $R_{H}$ and ionization radius $r_{\rm ion}$, respectively, for each $R$ considered. The adopted black hole masses are $M_1=M_2=10$\,M$_\odot$ and $M_{\bullet}=10^6$M$_\odot.$
\label{fig:mergertime}
}
\end{figure}

The evolution of binaries can be affected by interaction with the central SMBH as well as with field stars \citep{2009ApJ...690..795P,2012ApJ...757...27A,2014ApJ...782..101P}. Disruption by the SMBH, called \emph{ionization}, can occur for binaries with separations larger than the Hill radius $R_{H} = (2M_{\rm tot}/3M_{\bullet})^{1/3} R \approx 5\times10^{14}(R/10^{-2}\mbox{pc})$cm. This effect will only be constraining for the binaries at the smallest $R$ and greatest $r$ we consider (see Fig. \ref{fig:mergertime}). Close encounters with field stars will, on average, soften the binary until it is ionized if the binary's binding energy is smaller than the typical kinetic energy of field stars, $E_{k} \sim \langle M_{\star}\rangle \sigma^2$, with $\langle M_{\star}\rangle$ and $\sigma$ are the average mass and the one dimensional velocity dispersion of the stars, respectively. Approximating this dispersion with $\sigma^2\sim G M_{\bullet}/r$ \citep{2014ApJ...782..101P}, and assuming $\langle M_{\star}\rangle \sim M_{\rm tot}$, we find that binaries are eventually ionized for separations larger than $r_{\rm ion} \sim R M_{\rm tot}/M_{\bullet}$, where at the critical separation $r_{\rm ion}$ the binary's binding energy is equal to the typical kinetic energy of a field star \citep{2014ApJ...782..101P}.

We find that, for the assumed parameters, $r_{\rm ion}$ could limit the available radii for which dynamical friction dominates, especially for $r\lesssim 10^{-3}$pc (see Fig. \ref{fig:mergertime}). However, ionization may not be a hindrance if it occurs on a sufficiently long time scale \citep{2014ApJ...782..101P}. To estimate the ionization time scale $t_{\rm ion}$ of a binary, we consider a galactic nucleus within radius $R$ in which stellar-mass objects with mass $M_{\rm tot}$ make up $10\%$ of the mass of the central SMBH with mass $M_{\bullet}$. The number density of objects is therefore $n=0.1\,M_{\bullet}M_{\rm tot}^{-1}/(4\pi R^{3}/3)$. Estimating the cross section of a softening encounter to be $\pi r_{\rm ion}^2$, and taking the characteristic speed of field objects to be $\sigma$, the ionization time can be written as
\begin{equation}
t_{\rm ion} \approx \frac{1}{n \pi r_{\rm ion}^2 \sigma} \approx
10\times\frac{M_\bullet}{M_{\rm tot}}t_{\rm orb}\sim 10^{8}\,\text{yr}
\end{equation}
where $t_{\rm orb}$ is the orbital time for the BBH around the SMBH, and we obtained the numerical result for our fiducial parameters and $R=10^{-2}$\,pc. Similarly to \citet{2014ApJ...782..101P}, we conclude that the ionization time scale is long enough not to prevent mergers, unless the total mass of stars inside the orbit of the BBH is a significant fraction of the SMBH.

Ionization may also be important prior to the appearance of the AGN accretion disk, in determining the fraction of black holes residing in binaries. \cite{2016MNRAS.460.3494S} find that 60\%--80\% of stellar binaries are ionized after a Hubble time. The ionization fraction for heavier black hole binaries, however, may be much lower. This effect will reduce the total number of binary black holes available at the appearance of the AGN disk by less then a factor of 2.

% GAP OPENING RADIUS, both for and within the binary.
For a sufficiently massive binary, its gravitational torque pushes gas away from its orbit around the SMBH faster than it can be replenished by viscosity, and a gap is opened \citep{Duffell2015}. Following \citet{2011PhRvD..84b4032K} (their Eq. 44), we find for our adopted parameters that a gap will open for $R\gtrsim10^{-3}$pc, representing a significant fraction of the parameter space. Gap opening, nevertheless, does not mean that the amount of gas available for the binary needs to significantly decrease. For a single point-like low-mass perturber, hydrodynamical simulations have shown that even in the presence of a gap, the perturber can accrete gas from the background disk, through shocks occurring at the U-turns of horseshoe orbits inside the gap, at a rate comparable to the accretion rate in the unperturbed background disk \citep{1999ApJ...526.1001L,Dorazio+2016}.  Similar non-axisymmetric flows also develop across the gap, and can fuel a gas disk around the perturber, even when the perturber itself is a binary \citep{1999ApJ...526.1001L}. \cite{2011ApJ...726...28B} have explicitly computed the binary hardening rate within a thin gaseous disk, for a 15-15 ${\rm M_\odot}$ stellar binary orbiting a $3\times10^6{\rm M_\odot}$ SMBH, similar to the system envisioned here.  They found that binaries rapidly harden, despite the presence of a prominent gap in the disk, due to the formation of wound-up spiral wakes behind the stars in the minidisk. The merger time scale, based on the results of \citet{2011ApJ...726...28B}, scaled to our fiducial parameters at $R=10^{-2}$pc and $r=10^{14}$cm, is $t_{\rm merge}\sim3\times10^{5}$yr. This number is somewhat higher than our other two estimates based on \citet{2008ApJ...679L..33K} and \citet{2009ApJ...700.1952H}, although the long-term evolution of the stellar binary is not followed by \citet{2011ApJ...726...28B}. Most importantly, these results confirm the rapid merger of binaries, corroborating our argument. We note here that the above investigations of gap opening were carried out for disk parameters different from the our fiducial values. The simulation of gap opening in the thin disk scenario considered here, in the vicinity of the SMBH, is difficult and has not yet been carried out (although see \citep{DuffellMacFadyen2013} for a simulation of a disk whose thickness is only an order of magnitude larger). Such a simulation will be an important step in better understanding the role of gap opening. Nonetheless, the examination of the simulation scenarios indicates that gap opening is unlikely to qualitatively change the results presented here.

% MASS INCREASE

High accretion rate for the binary can lead to mass increase that can affect the observed binary mass distribution. With accretion efficiency $\epsilon=0.1$, at Eddington accretion, the black hole would increase its mass by only 3\% in 1\,Myr. For our rapid merger scenario, the increase therefore is not expected to be significant. However, as mentioned above, the rate at which the BBH is diverting gas from the AGN disk could be comparable to the accretion rate in this disk, and can therefore exceed the Eddington rate for the stellar-mass BHs by orders of magnitude (by a factor $\sim M_\bullet/M_{\rm tot}$).  We here assume that the BBH (and the minidisk in its vicinity) can only accept a small fraction of this fuel, and remain limited by the Eddington rate.  A higher gas supply rate could accelerate the BBH merger and lead to a rapid increase in the component BH masses.

% NOT ENOUGH GAS IS AVAILABLE TO CARRY AWAY THE ANGULAR MOMENTUM FROM THE BINARY
For our fiducial parameters, the total amount of gas within the binary's Hill radius is only about 0.1\% of the binary's mass, which implies a significant slow-down of the binary's orbital decay compared to the case when the mass of the circumbinary disk exceeds that of the binary. The migration rate we adopted includes this slow-down following the prescription in \citep{1995MNRAS.277..758S}. Note that the total amount of gas available within $R_{\rm disk}=10^{-2}$pc around the SMBH is greater than the binary's mass. The large influx of matter, given the accretion rate of the SMBH, ensures that gas is continuously replenished in the vicinity of the binary. For our fiducial parameters, gas with total mass of about 2500\,M$_{\odot}$ flows through a given radius within 10$^6$yr. This is $\sim 100$ times greater than the binary mass.

\section{Orbital alignment with disk}
\label{section:orbitalalignment}

As an accretion disk forms around a SMBH, typical stellar-mass binary black holes in the galactic nucleus are expected to orbit the SMBH with non-zero eccentricity and non-zero inclination with respect to the disk plane. These binaries are not immediately immersed in the disk to undergo rapid merger. These binaries, however, will periodically cross the accretion disk, interacting with it at every crossing. These interactions gradually erode the inclination between the plane of the BBH's orbit around the SMBH and the plane of the SMBH accretion disk, similarly to stars piercing AGN disks along their orbits \citep{1993ApJ...409..592A,2001A&A...376..686K,2005ApJ...619...30M}.

To estimate the number of binaries crossing the accretion disk of radius $R_{\rm disk}$, we assume that the orbital eccentricity of the CoM of the binary is drawn from an isotropic thermal distribution with $f(e)=\frac{1}{2}e$ and a strongly mass segregated number density $n(a)\propto a^{-2.5}$, which is also consistent with the massive O star distribution in the Galactic Center \citep{2009ApJ...697.1741B,2009ApJ...697.1861A}. Note that the cluster around $M_{\SMBH} \gtrsim 10^7$M$_\odot$ may not be relaxed within a Hubble time at least for a single-mass-component \citep{2013degn.book.....M}. However the core with massive components  assembles in a shorter time \citep{2009MNRAS.395.2127O}, which is less than a Hubble time for $M_{\SMBH} \lesssim 10^8$M$_\odot$. To obtain the total number of relevant binaries, we consider objects within the radius of influence $R_{\rm inf}$ of the SMBH (within which the SMBH determines the motion and dynamics of the objects; \citealt{2009MNRAS.395.2127O}). At the radius of influence we take the enclosed mass in stars to be $2M_{\SMBH}$ \citep{2012PhRvD..85l3005K}. Using $M_{\SMBH} = M_0 (\sigma/\sigma_0)^{k}$ according to the $M_{\SMBH}-\sigma$ relation, and taking
$\sigma_0 = 200\, \rm km/s$, $M_0=1.3\times 10^8\Msun$ and $k=4$ \citep{2008gady.book.....B,2009ApJ...698..198G}, we obtain
\begin{equation}\label{e:ri}
R_{\rm inf} = \frac{G M_{\SMBH}}{\sigma^2} = \frac{G M_0}{\sigma_0^2} \left( \frac{M_{\SMBH}}{M_0}\right)^{1-(2/k)} = 1.2\,M_6^{1/2}\,{\rm pc}\,,
\end{equation}
where $M_6\equiv M_{\SMBH}/10^6\Msun$.

We can express the fraction $f_{\rm cross}$ of objects within $R_{\rm inf}$ that will cross the SMBH accretion disk within $R_{\rm disk}$ to be
\begin{equation}\label{eq:fcross}
f_{\rm cross} = \iint\limits_{\shortstack{$a(1-e)<R_{\rm disk}$\\  $a<R_{\rm inf}$ } } n(a) f(e) \,4\pi a^2 \, da\, de.
\end{equation}
For $R_{\rm disk}\ll R_{\rm inf}$, this gives to leading order
\begin{equation}\label{eq:cross}
f_{\rm cross} = \frac56 \left(\frac{R_{\rm disk}}{R_{\rm inf}}\right)^{1/2}=0.075 \,M_6^{-1/4}\,,
\end{equation}
Evaluating Eq.~(\ref{eq:cross}) shows that $f_{\rm cross}=13\%$ ($2.4\%$) for $M_{\SMBH}=10^5\,\Msun$  ($10^8\,\Msun$).

To estimate the black-hole mass fraction within $R_{\rm inf}$, we take the results of \cite{2000ApJ...545..847M}. They considered a Salpeter stellar initial mass function, and assumed that stellar evolution will result in 0.6\,M$_\odot$ white dwarfs, 1.4\,M$_\odot$ neutron stars and 7\,M$_\odot$ black holes for the mass ranges 1--8\,M$_\odot$, 8--30\,M$_\odot$ and $30-100$\,M$_\odot$, respectively. They obtained a black hole mass fraction of 1.6\%. Assuming that black holes and neutron stars will dominate the central region due to mass segregation, this mass fraction rises to $\eta_{\rm bh} = 4\%$, which we will adopt in the following.

Another important parameter in our calculation is the fraction of black holes residing in binaries. While the binary fraction of massive stars is very large, some stellar binaries merge before forming two compact objects, they can be disrupted due to natal kicks at core collapse, or in some cases only one of the stars becomes a black hole. To estimate the fraction of black holes that will reside in binaries with other black holes, we neglect natal kicks as most black holes are either born through direct collapse without a supernova explosion, or receive small natal kicks that will not disrupt the binary \citep{2012ApJ...749...91F}. To count only those black holes whose companion will also be a black hole, we calculate the fraction of binaries for which both stellar progenitor has a mass $>30$\,M$_{\odot}$. With this choice we also neglect the possibility of binary merger prior to core collapse, which is more likely in the case of highly unequal mass binaries, such as black hole-neutron star progenitors \citep{2012ApJ...759...52D}. We take a Kroupa initial mass function with $\xi(M)\propto M^{-2.3}$ in the relevant mass range \citep{2001MNRAS.322..231K}, and adopt a uniform binary mass ratio \citep{2014ApJ...789..120B}. With these selections, the fraction of stellar progenitors with mass $30--100$\,M$_{\odot}$ whose companions also have masses in the range $30-100$\,M$_{\odot}$ is $\approx0.33$. With most massive stars residing in binaries, we adopt a black hole binary fraction $f_{\rm bin}=30\%$. We note here that in addition to binary evolution, this fraction may further increase due to dynamical binary formation channels in dense stellar environments.

With these parameters, the instantaneous number of BBHs crossing the disk is
\begin{equation}\label{e:Ncross}
N_{\rm cross} =  f_{\rm bin} f_{\rm cross} \frac{\eta_{\rm bh} M_{\SMBH}}{M_{\rm tot}} = 45\,\left(\frac{M_{\SMBH}}{10^6\mbox{M}_{\odot}}\right)^{3/4}\left(\frac{M_{\rm tot}}{10\mbox{M}_{\odot}}\right)^{-1}
\end{equation}
where all parameters are assumed to have the values in our fiducial model. BBHs which are initially not on disk-piercing orbits may later be scattered onto such orbits on the scalar resonant relaxation \citep{1996NewA....1..149R} timescale $t_{rr}=(M_{\SMBH}/M_{\rm tot})t_{\rm orb}$, where $t_{\rm orb}=2\pi a^{3/2} M_{\SMBH}^{-1/2}$G$^{-1/2}$. This is $0.5\,M_6^{1/2}\,$Myr at 0.01\,pc, less than the lifetime of bright AGN (10-100\,Myr). The number of black holes that cross and migrate into the accretion disk can therefore further grow.

We estimate the time scale of orbital alignment by comparing the change in the binary's perpendicular velocity $\Delta v_z$ with respect to the disk plane, to the total perpendicular velocity $v_z$. The binary's velocity will change each time it crosses the disk due to Bondi-Hoyle-Lyttleton accretion. For simplicity, we consider a single stellar-mass black hole with mass $M_{\rm tot}$ orbiting the central SMBH at radius $R$ on a circular orbit at inclination angle $\psi$ with respect to the disk, with orbital velocity $v_{\rm orb}=(GM_{\bullet} / R)^{1/2}$. Results are expected to be similar for a binary. Matter in the disk at $R$ will orbit at a comparable speed $v_{\rm orb}$. The relative velocity of the gas and the black
hole upon crossing is $\Delta v = v_{\rm orb}[(1 - \cos\psi)^{2} + \sin^2\psi]^{1/2}$ where $\psi$ is the inclination angle in the accretion disk plane. The duration of the crossing is $t_{\rm cross}\approx 2H/(v_{\rm orb}\sin \psi)$. Upon crossing, the black hole will accrete mass within its Bondi-Hoyle-Lyttleton radius $r_{\rm BHL} = 2 G M_{\rm tot}/(\Delta v^2 + c_{s}^{2})$. The accreted mass can be written as $\Delta M_{\rm cross} = \Delta v t_{\rm crossing} r_{\rm BHL}^{2} \pi \Sigma / (2 H)$, so for supersonic encounters $\Delta M_{\rm cross}\propto \Delta v^{-4} \Sigma $.
Note that $r_{\rm BHL}\approx (M_{\rm tot}/M_\bullet)R \ll R$, justifying the assumption of the formation and prompt accretion of a linear wake.

Since gas has, on average, zero velocity perpendicular to the disk, the velocity component of the black hole perpendicular to the disk will change upon crossing:
\begin{equation}
\frac{\Delta v_z}{v_z} = \frac{\Delta M_{\rm cross}}{M_{\rm tot}}
\end{equation}
Neglecting the mass increase of the black hole, the time scale of orbital alignment with the disk can be written as
\begin{equation}
\tau_{\rm align}\sim \frac{t_{\rm orb} v_z}{2 \Delta v_z}
\end{equation}
where $v_z = v_{\rm orb} \sin \psi$. The factor 2 is due to the 2 crossings per orbit. This alignment time scale is small for almost aligned orbits, and it is long for misaligned orbits (maximum  $\lesssim10^{9}$yr around $130^\circ$ for our fiducial parameters). Assuming uniform distribution in $\cos\psi$ between $-1$ and $1$ we find that for our fiducial model $(M_{\rm tot},M_{\SMBH})=(20,10^6)\Msun$, a fraction
$f_{\rm capt}\sim 2\%$, $\sim 6\%$, and $\sim20\%$ of the stars align with the disk within $10^6$, $10^7$, and $10^8$\,yr, respectively. These percentages are even higher for lower $M_{\SMBH}$ masses, e.g. for $10^5\,\Msun$, $(4, 14, 52)\%$ of the binaries get captured in less than $10^6$, $10^7$, and $10^8$ yr, respectively, and vice versa for higher masses.

This simplified calculation neglected several potentially important effects. One is orbital eccentricity. Eccentric orbits crossing the inner edge of the disk may have orbital periods longer by a factor $(R_{\inf}/R_{\rm disk})^{3/2}$, and $\Delta M_{\rm cross}$ decreases due to the larger velocity by a factor of $\lesssim 4$, further prolonging alignment. On the other hand, the alignment time scales with $v_{\rm orb}^{-4}\Sigma \propto R^{1.4}$ for orbits with a fixed orbital time, suggesting that the alignment is most efficient for circular orbits. We also neglected the effects of massive perturbers, which may efficiently excite the eccentricity and accelerate the rate of relaxation \citep{2007ApJ...656..709P}. Additionally, we neglected the effect of binary crossings on the disk. To check whether this assumption is justified, we computed the rate at which mass is removed from the disk by accretion onto the stellar-mass BHs during disk crossings.  We find that this removal rate (corresponding to the total accretion rate onto all disk-crossing stellar-mass BHs) is $\sim 10^5$ times lower than the accretion rate onto the SMBH.  We also estimated the heating of the disk due to (i) dynamical friction drag during disk crossings and (ii) absorption by the AGN disk of the radiation from accretion onto the stellar-mass BHs.  We find that these are $\sim 10^3$ and $\sim 10$ times lower than the local viscous heating rate for our fiducial disk-model. We conclude that the disc crossings do not significantly alter the mass or structure of the AGN disk.

Most importantly, stochastic torques from the spherical cluster cause chaotic orbital plane reorientation, a process called vector resonant relaxation \citep{1996NewA....1..149R,2015MNRAS.448.3265K}. If this process is much faster than alignment, alignment may be inhibited. On the other hand, vector resonant relaxation also reorients and warps the accretion disk \citep{2011MNRAS.412..187K}, which may change the piercing angle toward a favorable location helping alignment, if so, it may increase the encounter rates. The timescale for vector resonant relaxation is of order
\begin{equation}
t_{\rm vrr}=\frac{M_{\bullet}}{N^{1/2}M_\star} t_{\rm orb} = 2\times 10^5\,M_6^{0.31}\,{\rm yr}
\end{equation}
for the fiducial parameters ($0.01\,$pc), where $N$ is the enclosed number of stars and $M_\star$ is their RMS mass, and we assumed that the stellar cluster density profile follows $n\propto R^{-1.75}$ . Further out at the radius of influence, this timescale is of the order $10^7\,M_6^{3/4}$yr \citep{2006ApJ...645.1152H}.

There may be additional avenues for forming black hole binaries in accretion disks. Since the Bondi radius for our fiducial binary is $30\,R_{\odot}$, comparable to the radius of massive stars or red giants, binary stars may get captured in the disk, accrete from it and turn into BH binaries. Additionally, stellar binaries may be directly created in the outskirts of the accretion disk, grow into massive stars, get transported to the inner region by migration and become black holes there  \citep{GoodmanTan2004,Levin2007,McKernan+2012}.

We conclude that a significant fraction of stellar-mass black hole
binaries may align themselves with the accretion disk of the SMBH in a short time frame. In particular, the fraction of the $N_{\rm cross}$ disk-crossing BBHs that align with the disk by $f_{\rm capt}$ within an AGN lifetime, $\tau_{\rm align}=10^7\,$yr, is
$f_{\rm capt} = (14, 6, 3, 1, 0.1)\% $
for $M_{\SMBH}=10^{5,6,7,8,9}\,\Msun$, respectively. We emphasize that AGN disks may be expected to exist for even longer, $10^8$ years or more, with a reduced density accretion disk, which may still be sufficient to drive the BBH to merge. %Thus $\tau_{\rm align}=10^7\,$yr adopted here is conservative.

The value $f_{\rm capt}$ depends not only on $M_{\SMBH}$, but also on $M_{\rm tot}$. For BBH masses $M_{\rm tot}\lesssim 100$, we find the approximate scaling relation $f_{\rm capt}\propto M_{\rm tot}^{1/2}$. We will use this relation in the next section.

The total BBH merger rate in a single AGN is
\begin{equation}
\Gamma(M_{\SMBH}) = \frac{f_{\rm capt} N_{\rm cross}}{\tau_{\rm align}+\tau_{\rm merger}} = \frac{f_{\rm capt} f_{\rm bin} f_{\rm cross} \eta_{\rm bh} }{\tau_{\rm align} + \tau_{\rm merger}}\frac{M_{\SMBH}}{M_{\rm tot}}
\end{equation}
where we have substituted Eq.~(\ref{e:Ncross}). This expression holds
with our fiducial numbers until the BBHs on crossing orbits are all
removed from the cluster. If this happens within the vector resonant
relaxation timescale, the merger rate drops to zero until vector
resonant relaxation (VRR) can reshuffle the orbits.
However this never happens for our fiducial model parameters, which give
%$\Gamma = (2\times 10^{-8}, 10^{-7}, 1.6\times 10^{-5}, 5\times 10^{-7}, 1.5\times 10^{-5})\,{\rm yr}^{-1}$ for $M_{\SMBH}=10^{5,6,7,8,9}\,\Msun$, respectively.
$\Gamma = (10^{-7}, 2\times10^{-7}, 7\times 10^{-7}, 1.3\times 10^{-6}, 7\times 10^{-7})\,{\rm yr}^{-1}$ for $M_{\SMBH}=10^{5,6,7,8,9}\,\Msun$, respectively.

\subsection{Stellar dynamics prior to black hole formation}

The above description presumes that black hole binaries are initially present outside of the accretion disk of the active galaxy, implying that the black holes' progenitor stars did not significantly interact with, nor were formed within, the accretion disk. Regarding the latter case, for larger radii from the SMBH, the self-gravitating accretion disk can become fragmented, inducing star formation within the disk. This scenario can result in similar rapid merger discussed above if the formed stars have enough time to collapse and produce BBHs before merging themselves. We further refer the reader to \cite{2016arXiv160204226S}, who discuss in detail the possibility of binary formation within the accretion disk at $\gtrsim 0.1$\,pc, outside of the region of interest discussed here.

Massive stars present within galactic nuclei could migrate into the accretion disk similarly to black holes. However, due to their very short lifetime ($O(10^6\mbox{yr})$), only those stars will undergo this migration that are formed within the galactic nucleus, and during the time in which the galaxy was active. Black holes that were formed farther from the center and migrated in due to mass segregation are therefore not affected by the progenitor stars' interaction with the accretion disk. Also, black holes that were formed prior to the onset of the SMBH accretion are not affected either. We therefore do not expect stellar interaction with the SMBH accretion disk to significantly affect our expected rate or dynamics.

\section{Rate within LIGO's horizon}
\label{section:RATE}

To estimate the rate of binary black hole mergers within AGN accretion disks, we take into account the distribution of SMBH and BH masses. We use the lognormal fit to the observed AGN mass function in the local universe \citep{2007ApJ...667..131G,2009ApJ...704.1743G}
\begin{equation}
\frac{\mathrm{d} n_{\rm AGN}}{\mathrm{d} M_{\SMBH}} = \frac{3.4\times 10^{-5}{\rm Mpc}^{-3}}{M_{\SMBH}}\times 10^{- \left[\log (M_{\SMBH}/\Msun) - 6.7\right]^2/0.61}
\end{equation}
For simplicity, we consider only equal-mass binary black holes. We take the number density of binaries to be $p(M_{\rm tot})\propto M_{\rm tot}^{-2.5}$ with minimum and maximum masses of 10\,M$_{\odot}$ and 100\,M$_{\odot}$, respectively \citep{2016arXiv160604856T}. This mass distribution will modify $N_{\rm cross}$ (see Eq. \ref{e:Ncross}). The number density of BBHs with masses $M_{\rm tot}$ will be
\begin{equation}
\frac{\mathrm{d}{N_{\rm cross}}}{\mathrm{d}{M_{\rm tot}}} = 2.3 f_{\rm bin} f_{\rm cross} \eta_{\rm bh} M_{\SMBH}M_{\rm tot}^{-3.5}
\end{equation}
where the factor 2.3 comes from the normalization of $p(M_{\rm tot})$. The corresponding BBH merger rate density in a single AGN is
\begin{equation}
\frac{\mathrm{d}{\Gamma}}{\mathrm{d}{M_{\rm tot}}} = \frac{f_{\rm capt}(M_{\rm tot})}{\tau_{\rm align}+\tau_{\rm merger}}\frac{\mathrm{d}{N_{\rm cross}}}{\mathrm{d}{M_{\rm tot}}}
\end{equation}
The total merger rate density from all AGN over all BBH masses is
\begin{equation}
\mathcal{R}= \iint \frac{\mathrm{d} n_{\rm AGN}}{\mathrm{d} M_{\SMBH}} \frac{\mathrm{d}{\Gamma}}{\mathrm{d}{M_{\rm tot}}} \mathrm{d}{M_{\SMBH}} \mathrm{d}{M_{\rm tot}} \approx 1.2\, {\rm Gpc}^{-3} {\rm yr}^{-1}\,.
\end{equation}

We define the LIGO horizon distance as the luminosity distance at which the gravitational-wave signal-to-noise ratio of an optimally oriented source at an optimal sky location is $\rho=8$ for a single detector \citep{2010CQGra..27q3001A}. We adopt a horizon distance of $D_{h} = 450$\,Mpc\,$(M_{\rm tot}/2.8)^{5/6}$ for Advanced LIGO at design sensitivity \citep{2015ApJ...806..263D}, which provides a good estimate for the relevant mass range, given that the binary is equal-mass. Averaging over direction and orientation decreases this effective distance by a factor of 2.26 \citep{2013CQGra..30l3001B}. The corresponding comoving volume is $V_{c}(M_{\rm tot}) = 4/3 \pi (D_{h}/2.26)^3(1+z)^{-3}$, where $z$ is the redshift corresponding to $D_{h}/2.26$ luminosity distance. Neglecting the merger rate evolution with redshift, we find an expected detection rate for these types of sources is
\begin{equation}
\Gamma_{\rm aLIGO}=\int V_c(M_{\rm tot}) \frac{\mathcal{\mathrm{d}R}}{\mathrm{d}M_{\rm tot}} \mathrm{d}M_{\rm tot} \approx 13\,{\rm yr}^{-1}
\end{equation}
%With the sensitivity level of Advanced LIGO in its first observation period (O1), we characterize the BBH horizon distance to be a factor 3 less than at design sensitivity, which corresponds to an expected number of detections of $\sim 0.1$.
While this estimate is subject to uncertainties due to the approximations we adopted above, it is clear that detecting gravitational waves from the merger of massive stellar binaries with Advanced LIGO may be feasible. More detailed modeling is needed to understand the implications of potential detections or nondetections.

\section{Electromagnetic signature}
\label{section:EM}

The inspiral and merger of binary black holes within a gaseous medium can be accompanied by luminous electromagnetic radiation.  In a uniform gas cloud, luminosity can be enhanced compared to a single black hole with the same mass, by up to several orders of magnitude near the merger, due to shocks produced by the orbital motion of the binary \citep{2010PhRvD..81h4008F}. This uniform gas approximation, nevertheless, omits the effect of increased gas pressure due to the rotating AGN accretion disk, which can decrease total accretion. A configuration similar to the one discussed in this paper, \cite{2016arXiv160204226S} also find that accretion to the binary can exceed the Eddington rate by orders of magnitude.

Such super-Eddington accretion onto 10--100\,M$_\odot$ black holes can produce a fast, bright (super-Eddington) electromagnetic transients through the disk's thermal emission \cite{2014ApJ...796..106J,McKinney+2014,2016ApJ...822L...9M}, and possibly through driving relativistic outflows \citep{2006ARA&A..44...49R,2005A&A...429..267B}.

For binaries buried within the accretion disk, emission can be reprocessed as it passes through optically thick disk. This can convert high-energy emission to optical/infrared, and spread out the emission in time. However, as we saw in Section \ref{sec:regionofvalidity}, most of the binaries will open gaps within the AGN disk, enabling their emission to leave the AGN without needing to pass through the disk.

To characterize the detectability of high-energy emission from the accretion binary, let the bolometric luminosity of the binary shortly before the merger be $L_{\rm bol} = \eta_{bol}L_{\rm Edd}$, and assume that a fraction $\epsilon_{\gamma}$ of this luminosity is radiated in the gamma-ray
band. The observed gamma-ray flux at Earth from a binary at luminosity distance $D_{L}$ can then be written as
\begin{equation}
\Phi_{\gamma} = 10^{-14} \,\eta_{bol} F\left[\frac{M_{\rm tot}}{100\,\text{M}_{\odot}}\right]
\left[\frac{D_{L}}{100\,\text{Mpc}}\right]^{-2}\frac{\text{erg}}{\text{cm}^{2}\text{s}}
\end{equation}

The Fermi-LAT telescope is sensitive to gamma-rays above $\sim
100$\,MeV, with source sensitivity
$\sim10^{-6}$phot\,cm$^{-2}$s$^{-1}$ for $\sim1$\,hr exposure
\citep{2009ApJ...697.1071A}. Estimating this sensitivity limit with
$\Phi_{\rm LAT}\sim
10^{-4}$MeV\,cm$^{-2}$s$^{-1}\approx10^{-10}$erg\,cm$^{-2}$s$^{-1}$,
we find that the detection of a binary black hole merger requires
super-Eddington luminosity with $\eta_{bol}\gtrsim10^{4}$,
where we assumed $\epsilon_\gamma \sim 1$ and favorable source parameters.
Although the studies mentioned above have shown that luminosities as
high as $10-100~{\rm L_{\rm Edd}}$ can be achieved, the detection of a
gamma-ray counterpart would require a further increase by at least two
orders of magnitude above this value.

Alternatively, the source's effective luminosity can further increase if the large accretion rate results in the production of a beamed, relativistic outflow, in which internal dissipation results in non-thermal, high-energy emission. For a beaming factor $\mathcal{O}(100)$, comparable to that observed in gamma-ray bursts \cite{2013CQGra..30l3001B}, detection prospects can significantly increase even for $10-100~{\rm L_{\rm Edd}}$ emission.

If the binary produces a comparable luminosity in X-rays, it could be
observable with X-ray detectors, such as the Chandra X-ray observatory
\citep{1999astro.ph.12097W}. With a point-source sensitivity of
$4\times10^{-15}$erg\,cm$^{-2}$s$^{-1}$ between 0.4-6\,keV for
$10^{4}$s exposure, Chandra could detect a binary with $\eta_{bol}\sim
1$ and $\epsilon_{X}=0.1$ out to $100$\,Mpc. Since Chandra's field of view is very small compared to the localization uncertainty of LIGO-Virgo, additional directional constraints are needed from either a catalog of plausible host galaxies \citep{2016ApJ...826L..13A,2015ApJ...801L...1B,2016ApJ...816...61B},  or other cosmic messengers with all-sky detection capabilities, such as gamma rays \citep{2016ApJ...826L..13A,2016ApJ...826L...6C} and high-energy neutrinos \citep{2016PhRvD..93l2010A,2014PhRvD..90j2002A,2012PhRvD..85j3004B}.
Alternatively, this scenario can be interesting because of the
potential of observing a binary in an AGN nucleus that has not merged
yet, allowing for significantly longer observation window. Such a discovery would be informative of, e.g., the binary population
within accretion disks.  The proposed X-ray surveyor satellite, with
30-100 times Chandra's throughput, and a significantly larger field of
view, could be used to search for EM emission from similar BBH GW
sources in the future \citep{XRS2015}.

The above estimates show that emission from binaries within AGN accretion disk may be a promising electromagnetic counterpart for binary black hole mergers. As key open questions remain, further numerical and theoretical investigations will be critical in better understanding the accretion and emission processes and detection prospects.

\section{Conclusion}
\label{section:conclusion}

We examined the fate of stellar-mass binary black holes within an
active galactic nucleus. We found that a significant fraction of the
binaries migrates into the accretion disk around the supermassive
black hole in the center of the active galaxy. Within the accretion
disk, we found that binaries rapidly merge ($\lesssim 1$\,Myr) via
dynamical friction and, at the end stage, gravitational
radiation. This scenario presents an interesting opportunity for
gravitational wave observations of binary black holes within a
high-density circumbinary medium. We estimate that the detection rate of gas-induced mergers observable with Advanced LIGO at design sensitivity is $\sim 20$\,yr$^{-1}$. This corresponds to $\sim20$\% chance that such an event is detected during Advanced LIGO's first observation period O1.

We discussed the prospects of detecting the electromagnetic counterparts of these binary black hole
mergers. We found that detectable radiation at the time of the merger
would require a highly super-Eddington luminosity, which is unlikely,
although not excluded, as the relevant emission physics is
highly uncertain. Two interesting observation scenarios are (i) if
high accretion rate gives rise to energetic outflows that produce
non-thermal, high-energy emission within the outflow; (ii) if one
searches for nearby binaries that are within the accretion disk of an
AGN, but is not yet close to merging.

To better understand the gas-induced merger channel proposed here, as well as its prospects of producing an observable electromagnetic counterpart, it will be important to (i) carry out detailed simulations of binary black hole accretion close to merger, incorporating realistic AGN disk properties and gap opening by the binary; (ii) study the electromagnetic emission of such scenarios, in particular whether a relativistic jet can be driven by the system; (iii) carry out a similar analysis for less dense accretion disks, and other potentially gaseous environments, e.g., in globular clusters.

\acknowledgements

The authors want to thank Cole Miller for useful suggestions. IB and SM are thankful for the generous support of Columbia University in the City of New York and the National Science Foundation under cooperative agreement PHY-1447182. ZH acknowledges support from NASA ATP grants NNX11AE05G and NNX15AB19G and from a Simons Fellowship in Theoretical Physics. This work was supported in part by the European Research Council under the European Union's Horizon 2020 Programme, ERC-2014-STG grant GalNUC 638435, and it was completed in part [by BK] in the Aspen Center for Physics, which is supported by NSF grant \#PHY-1066293.

%#################################################################
%\bibliographystyle{h-physrev}
\bibliographystyle{yahapj}
%\bibliography{Refs}

\end{document}